\documentstyle[11pt,twoside,pasp3D,epsf]{article}

\begin{document}

\title{Focussing Image Slicers: Refractive and Reflective}

\author{E.H. Richardson}
\affil{EHR Optical Systems, 1871 Elmhurst Pl.,Victoria, V8N 1R1, Canada}

\author{A. Moore}
\affil{CREO, 3700 Gilmore Way, Vancouver, V5G 4M1, Canada}
\author{T. Tilleman}
\affil{NOAO, 950 N. Cherry Ave., Tucson, AZ, 85726-6732}
\author{David Crampton}
\affil{National Research Council of Canada, Victoria, V8X 4M6, Canada}

\begin{abstract}
A number of design options for image slicers for NGST and Gemini are
being investigated.  These image slicers are all of the focussing type
and both refractive and reflective solutions are being explored. One
such device, an image slicer that focuses 10 slices on a
spectrograph slit is now in operation at the McMath Solar telescope. It
consists of three lenslet arrays, and additionally acts as a focal
reducer and provides correction for astigmatism of the telescope. A
combined refractive and reflective slicer designed for use on NGST
delivers near-diffraction limited images for up to 40 slices.
\end{abstract}

\section{Introduction}

Image slicers were initially designed and used to reformat the seeing
disk of stars into a shape that could efficiently be introduced into
the slit of a spectrograph. The first of these, designed by Bowen
(1938), reflected nearly-focussed starlight parallel to the slit and
then reflected slices of it into the slit using thin mirror facets
whose width equalled the slit width. An inherent disadvantage of this
type of slicer is that only one slice is in focus at the slit, although
it, and the more elegant Walraven image slicer design, were still
useful for their original purpose. In practice, the non-uniform
illumination of the spectrograph collimator frequently led to
substantial problems and they were not extensively used. The
superpositioning image slicers (Richardson, 1968) were designed for use
with a single-slit spectrograph in which photographic plates were used
as detectors.  Each slice of the blurry star is elongated and then
superimposed on the slit providing deliberate scrambling (for more
accurate radial velocities and more uniform illumination along the
slit).  In this way, the image slicer increased slit transmission and
improved the spectrophotometric accuracy, especially with non-linear
detectors such as photographic emulsions. A more detailed discussion of
these image slicers is given by Richardson, Fletcher \& Grundmann
(1984).

Now that superb spatial resolution can be achieved through use of
adaptive optics or by telescopes in space, image slicers are being
considered to reformat the more-or-less circular images of resolved
targets such as galaxies into long slit-shaped apertures in order to
obtain 2-D spectral information or to study kinematics of the whole
object at once. Of course, to preserve the spatial information, these
new designs (e.g., Diego 1994; Content 1997)  must ensure that the
slices are in proper focus. We are examining both all-reflective and
all-refractive types of these focussing image slicers. In the following
sections we describe two specific examples of these Richardson
Focussing Image Slicers (RFIS).

\section{RFIS 2.1: McMath Focussing Image Slicer}

As mentioned above, a disadvantage of the Bowen-Walraven type of image
slicer is that only one slice is in perfect focus on the slit, the
others being out of focus because the light travels a larger (or
smaller) distance before reaching the slit. In 1994, one of us (EHR)
designed a stacking type image slicer for Drew Potter of Lockheed/NASA,
Houston for use on the McMath solar telescope spectrograph.  The basic
goal was to slice the planet Mercury into 10 slices and then stack these
along the 20 mm long slit, demagnified by a factor of 2, with each slice in focus
in order to preserve the spatial resolution of planetary features along
the slit.

The detailed specifications were:
\begin{itemize}
\item slice the 10\arcsec\ disk of the planet Mercury into ten 4mm $\times$ 0.4mm  slices
\item demagnify the slices by a factor of 2 (from f/54 to f/27) 
\item correct for astigmatism of telescope
\item maintain focus of slices along the slit 
\item conform to mechanical constraints of the existing Bowen/Walraven image slicer
\end{itemize}

The optical design, by EHR Optical Systems, comprises 3 sets of lenslet
arrays, 30 lenslets total, all BK7 glass. The lens fabrication was
carried out by Lumonics (formerly Interoptics) Ottawa, and the
mechanical assembly was designed by A. Moore. An overall layout and a
photograph of the completed slicer are shown in Figure~\ref{RFIS2}. The
components were successfully aligned and installed in 1996. Performance
tests are ongoing, but it apparently works as expected.

\section{Richardson Focussing Image Slicers designed for NGST}

\subsection{RFIS 3.1}
An all-refractive image slicer, similar to McMath design described
above, was designed for NGST. In this case, however, the design is
considerably simpler since it doesn't have to incorporate focal
reduction nor to compensate for telescope aberration. A ten-slice image
slicer with an entrance aperture of 3\arcsec $\times$ 1\arcsec (ten
0\farcs1 slices) is straightforward and larger ones appear possible.

\subsection{RFIS 3.2}

\begin{figure}
\plottwo{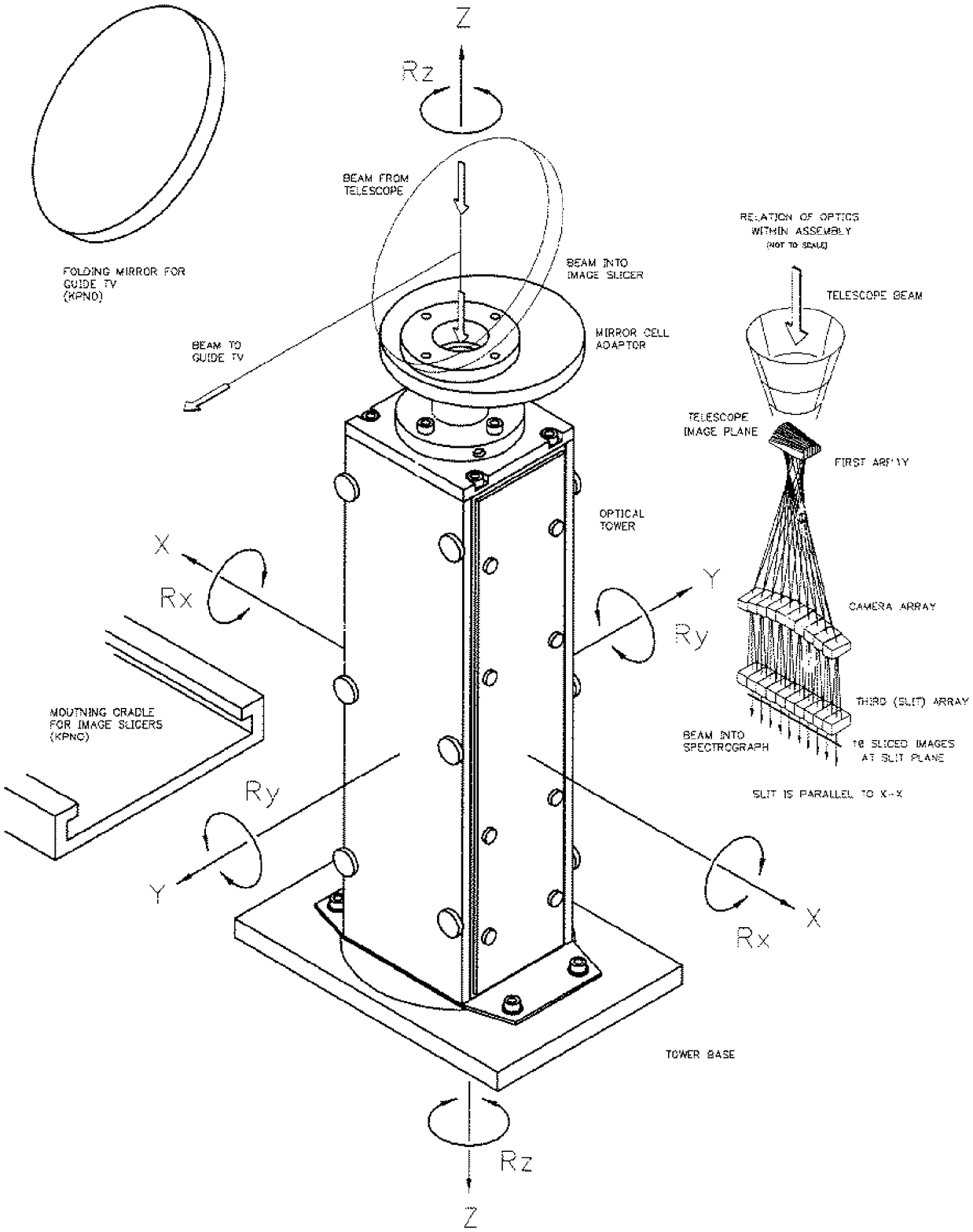}{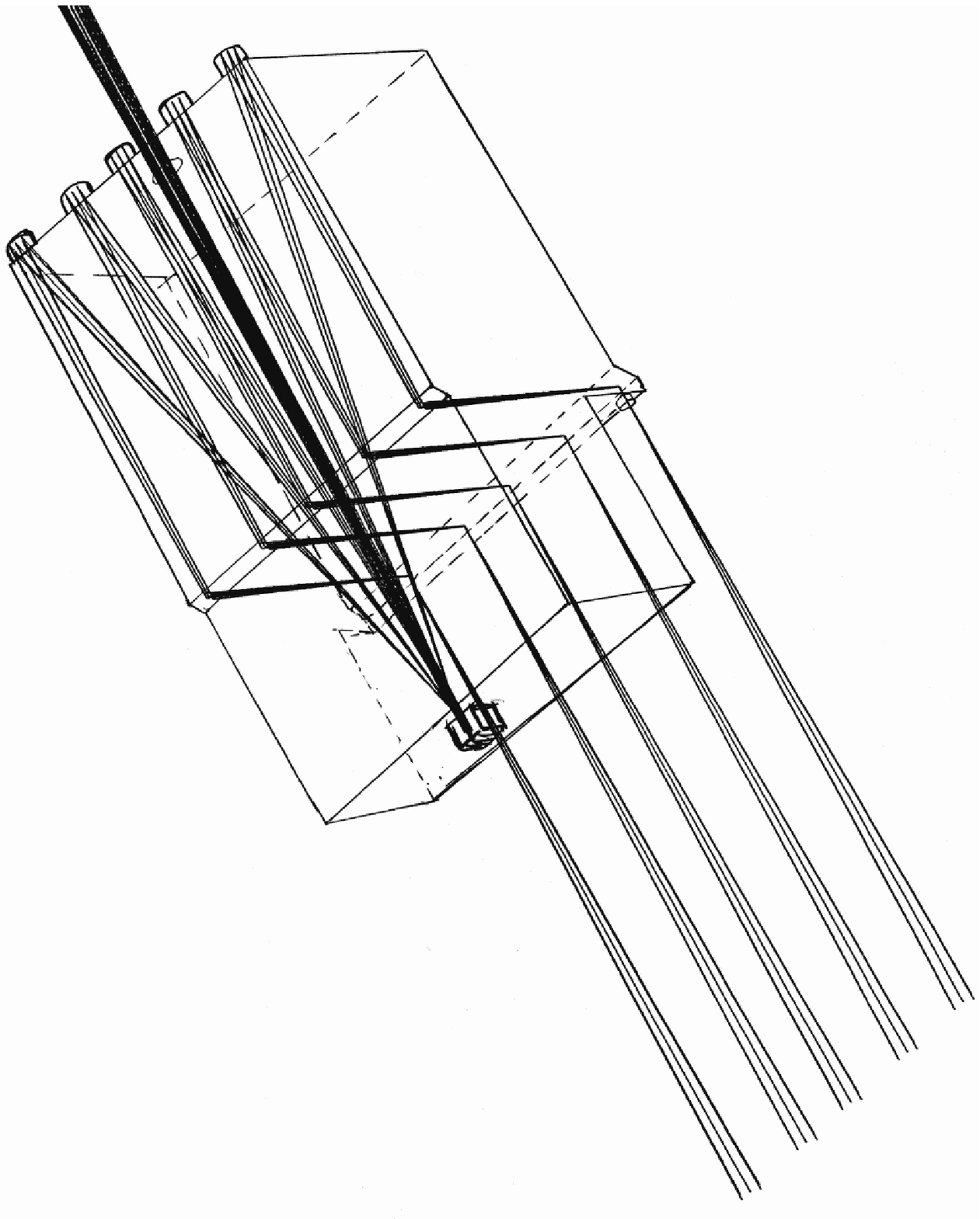}
\caption{Left: The overall mechanical layout of RFIS 2.1. Three sets of
lenslets arrays are used to slice up the image and refocus the slices
into the slit. A beam splitter above the entrance to the device is used
to deflect some light to the guide camera. Right: A schematic diagram
of RFIS 3.2. Light from the telescope enters the glass block from the
upper left, reflects off the first mirrorlet array, returns to the
top and reflects off another (long) array, and then is redirected by
two nearly right angle prisms. It then passes through an array of field
lenses just before the slit and into the spectrograph. The longest
dimension is 126mm.}
\label{RFIS2}
\end{figure}

A combined refractive and reflective design appears to offer more
advantages.  This design, shown on the right in Figure~\ref{RFIS2},
uses concave back-surface mirrors (the incident and reflected light
beams are inside glass) and has only two air-glass surfaces (one at the
2\farcs0 $\times$ 3\farcs0 aperture, the other at the 65\arcsec\ long
slit). The powered optics consist of the first mirrorlet array, the
reimaging mirrorlet array and the field lenslet array located at the
exit slit. Two near-right angle internally reflecting prisms redirect
light to the fixed slit location. All of the optical surfaces are
spherical or flat.

Features of the prototype design:
\begin{itemize}
\item Slices a 2\arcsec $\times$ 3\arcsec\ object at NGST focus into 20 0\farcs1 slices 
\item Slices form a slit 65\arcsec (68mm) long  times 0\farcs1 wide.
Each slice is offset from the next by 0\farcs12 along the slit to preserve spatial
information at the ends of the slices.
\item Input is offset laterally from output by 43\arcsec, but location of the telescope focus/spectrograph input focus is preserved
\item RMS image diameters are typically 46 microns (0\farcs05) i.e., nearly diffraction-limited. The worst image (outermost) is 57 microns or 0\farcs06
\item Very stable alignment since mirrors and lenses are bonded to a block of glass of the same material. For the NGST core wavelength range (1 $-$ 5 $\micron$), CaF$_2$ would be used.
\item Good focus maintained over entire (1 - 5 $\micron$) wavelength range 
\item Very high throughput, very low light loss and very low scattered
light since there are only two air/glass surfaces and since all other
reflective surfaces are internal.
\item Design compensates for non-telecentricity of NGST telescope design.
\item Very compact (126mm by about 12mm wide)
\end{itemize}

This design can be extended to provide either longer slices or to
provide more slices of the same length. If one wished to sample the
diffraction-limited images at 2$\micron$ with two pixels, then the
nominal 4K detector would allow a total slit length equivalent to
100\arcsec. Hence a slightly larger format is possible with the
strawman detector. Obviously, if one chose to accept lower spatial
resolution then the field could be substantially enlarged. Another
option to increase the field size (while retaining the superb resolution)
would be to increase the length of
the detector mosaic. 

The McMath image slicer demonstrates that such an
image slicer could be easily achieved in practice.  The lenslets for
the McMath device were made by fabricating two lenses for each stage
which were then cut up (every second lenslet was destroyed by the saw
cuts, so two lenses were required) to produce the lenslets.  Subsequent
alignment of the lenslets and of the image slicer posed no difficulties.
Hence, the RFIS 3.2 design concept appears to provide an excellent
image slicer for NGST scientific applications.

\acknowledgments

We are grateful to Murray Fletcher for assistance with the diagram of
RFIS 3.2 which is shown in Figure 1.


\begin{references}

\reference Bowen, I.S. 1938, Ap. J., 88, 113
\reference Content, R. 1997, SPIE 2871, 1295
\reference Diego, F. 1994, SPIE 2198, 525
\reference Richardson, E.H. 1968, JRASC, 62, 313
\reference Richardson, E.H., Fletcher, J.M., \& Grundmann, W.A. 1984, Proc. IAU Colloq. No. 79: Very Large Telescopes, their Instrumentation and Programs, Garching, 469
 
\end{references}
\end{document}